\documentclass[aps,prb,twocolumn,showpacs,superscriptaddress]{revtex4-1}
\usepackage{graphicx}
\usepackage{array}
\usepackage{amssymb}
\usepackage{amsfonts}
\usepackage{amsmath}
\usepackage{mathrsfs}
\usepackage{color}
\usepackage{booktabs}
\usepackage{threeparttable}
\usepackage{multirow}
\usepackage{subfigure}
\usepackage{epsfig}
\usepackage{threeparttable}
\usepackage{chngpage}
\usepackage{float}
\usepackage{color}
\usepackage{bm}
\usepackage{graphicx}
\usepackage[toc]{appendix}

\makeatletter
\renewcommand\@biblabel[1]{#1.}
\makeatother

\begin{document}

\title{Robustness of persistent currents in two-dimensional Dirac
systems with disorders}

\author{Lei Ying}
\affiliation{School of Electrical, Computer, and Energy Engineering,
Arizona State University, Tempe, AZ 85287, USA}

\author{Ying-Cheng Lai} \email{Ying-Cheng.Lai@asu.edu}
\affiliation{School of Electrical, Computer, and Energy Engineering,
Arizona State University, Tempe, AZ 85287, USA}
\affiliation{Department of Physics, Arizona State University, Tempe,
AZ 85287, USA}

\date{\today}

\begin{abstract}

We consider two-dimensional (2D) Dirac quantum ring systems formed by
the infinite mass constraint. When an Aharonov-Bohm magnetic flux is
present, e.g., through the center of the ring domain, persistent
currents, i.e., permanent currents without dissipation, can arise.
In real materials, impurities and defects are inevitable, raising the
issue of robustness of the persistent currents. Using
localized random potential to simulate the disorders, we investigate
how the ensemble averaged current magnitude varies with the disorder
density. For comparison, we study the nonrelativistic quantum counterpart
by analyzing the solutions of the Schr\"{o}dinger equation under the same
geometrical and disorder settings. We find that, for the Dirac ring
system, as the disorder density is systematically increased, the average
current decreases slowly initially and then plateaus at a finite nonzero
value, indicating remarkable robustness of the persistent currents. The
physical mechanism responsible for the robustness is the emergence of
a class of boundary states - whispering gallery modes. In contrast, in
the Schr\"{o}dinger ring system, such boundary states cannot form and the
currents diminish rapidly to zero with increase in the disorder
density. We develop a physical theory based on a quasi one-dimensional
approximation to understand the striking contrast in the behaviors of the
persistent currents in the Dirac and Schr\"{o}dinger rings. Our 2D Dirac
ring systems with disorders can be experimentally realized, e.g., on the
surface of a topological insulator with natural or deliberately added
impurities from the fabrication process.

\end{abstract}
\maketitle

\section{Introduction} \label{sec:intro}

Persistent or permanent currents, i.e., currents requiring no
external voltage with zero resistance, were traditionally thought to
occur only in superconductors. However, about three decades ago, it was
theoretically predicted~\cite{BIL:1983} that such dissipationless currents
can emerge in normal metallic or semiconductor ring systems subject to a
central Aharonov-Bohm (AB)~\cite{AB:1959} magnetic flux. In particular, if 
the ring size is smaller than the elastic scattering length, the electron 
motion in the entire domain will become ballistic, effectively eliminating 
scattering. If the ring size is larger than the elastic 
scattering length, the electron's behavior will be diffusive with a 
current proportional to $1/\tau_D$, where $\tau_D$ is the diffusion
time around the ring. While the environmental temperature
needs to be sufficiently low to reduce inelastic scattering from
phonon-electron and/or electron-electron interactions for the currents to be
observed~\cite{BIL:1983,CRG:1989,Schmid:1991,Bouchiat:2008}, the metallic
material itself remains ``normal'' (i.e., not superconducting). The
remarkable phenomenon of persistent currents was subsequently observed
experimentally in a large variety of settings~\cite{LDDB:1990,CWBKGK:1991,
MCB:1993,RSMHE:2001,KBFGG:2007,BSPGVGH:2009,BKBHM:2009,CNSJH:2013}, all
being nonrelativistic quantum systems.

Persistent currents in nonrelativistic quantum systems are
vulnerable to material impurities, which fundamentally restricts the phenomenon
to systems at or below the mesoscopic scale. Indeed, in real materials
disorders are inevitable, which can dramatically reduce the phase coherent
length due to enhanced random scattering. In general, random disorders
can remove the energy degeneracies and induce level repulsion, opening energy
gaps and destroying the conducting state. As a result, disorders in metallic
or semiconductor systems, 1D or 2D, tend to diminish the persistent
currents~\cite{CGRS:1988,VR:1991,WKK:1994,CP:1995,PP:2005,BL:2008,BEI:2010}.
As the strength of the disorder is increased, the currents
decay exponentially to zero~\cite{CGRS:1988,BEI:2010}.

Recent years have witnessed a tremendous development and growth of
interest in 2D Dirac materials such as graphene~\cite{Novoselovetal:2004,
Bergeretal:2004,Novoselovetal:2005,ZTSK:2005,Netoetal:2009,Peres:2010,
SAHR:2011}, topological insulators~\cite{TIs:RMP}, molybdenum disulfide
(MoS$_2$)~\cite{RRBGK:2011,WKKCS:2012},
HITP [Ni$_3$(HITP)$_2$]~\cite{Sheberlaetal:2014},
and topological Dirac semimetals~\cite{YZTKMR:2012,Liuetal1:2014,
Liuetal2:2014,YK:2015,JJY:2015}.
The quantum physics of these 2D materials is governed by the Dirac
equation or the generalized Dirac-Weyl equation~\cite{XL:2016,XL:2017},
and there were studies of persistent currents, e.g., in
graphene~\cite{PC:Grap,PC:Grap1,PC:Grap2,PC:Grap3,PC:Grap4,PC:Grap5,
PC:Grap6,PC:Grap7,PC:Grap9,PC:Grap10} and other Dirac
materials~\cite{PC:TI,PC:DF}. The effects of boundary deformation on
the persistent currents were recently investigated~\cite{XHLG:2015,XHL:2016},
where it was found that, even when the deformation is so severe that
the corresponding classical dynamics in the 2D domain becomes fully
chaotic, persistent currents can sustain. The physical origin of the so-called
superpersistent currents~\cite{XHLG:2015,XHL:2016} can be attributed to the
emergence and robustness of a type of quantum states near the boundaries of
the domain, which carry a large angular momentum and correspond essentially
to the whispering gallery modes (WGMs) that arise commonly in optical
systems~\cite{NSCGC:1996,GCNNSFSC:1998,SKV:2002,Vahala:2003} and can occur
in nonrelativistic quantum electronic systems~\cite{WGM:electrons} as well.
The Dirac WGMs are insensitive to boundary deformations, which
may be intuitively understood as a consequence of the zero flux boundary
condition required for nontrivial, physically meaningful solutions of the
Dirac equation. In spite of these efforts, the effects of bulk disorders
on persistent currents in 2D Dirac systems remains to be an open issue. In
particular, since there are random scattering sources inside the domain
with a finite probability of occurrence even near the boundary, it is not
intuitively clear whether the Dirac WGMs and hence persistent currents
can still exist when there are random impurities in the ring domain.

\begin{figure}
\centering
\includegraphics[width=\linewidth]{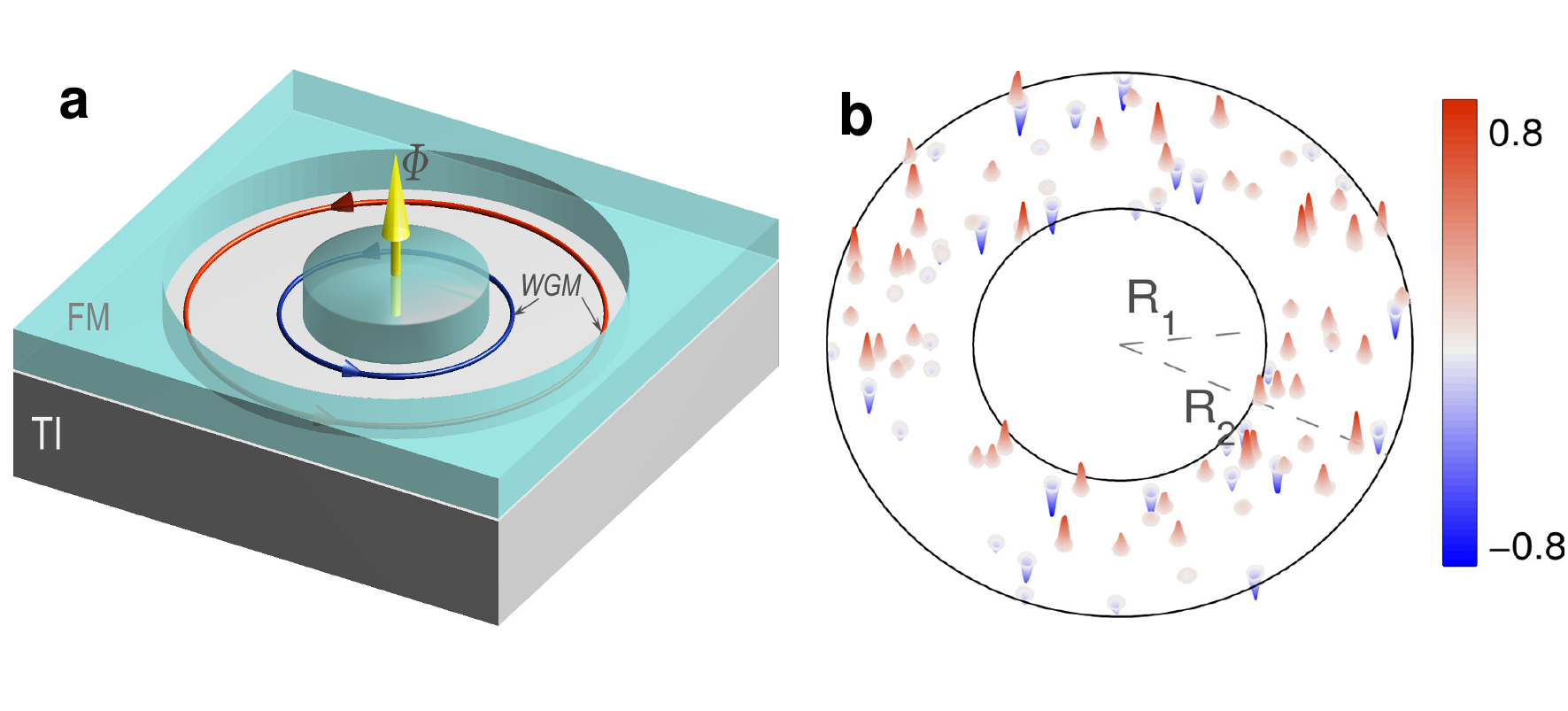}
\caption{ (a) Schematic illustration of a ring domain with an
AB magnetic flux through the center. The light blue color denotes the
regions of infinite mass. Red and blue loops illustrate the eigenstates
near the outer and inner boundaries, respectively. (b) Schematic
illustration of random disorders uniformly distributed in the
ring region, with their strength denoted with different colors.
Experimentally a Dirac ring can be generated by placing a ferromagnetic
insulator of proper shape on the surface of a topological
insulator~\cite{XHLG:2015,XHL:2016}.}
\label{fig:schematic}
\end{figure}

In this paper, we investigate the effects of random disorders on persistent
currents in 2D relativistic quantum systems. To be concrete, we consider a
Dirac ring domain with a vertical magnetic flux through the center, as
shown in Fig.~\ref{fig:schematic}(a). To completely constrain a Dirac fermion
within the domain, we impose the infinite mass boundary condition originally
introduced by Berry~\cite{BFM:1987} into the study of chaotic neutrino
billiard, which is experimentally realizable through a proper arrangement of
ferromagnetic insulation~\cite{WKASJHM:2013}. We assume uncorrelated disorders
throughout the domain, which can be simulated using localized, random electric
potential uniformly distributed in the domain, as illustrated in
Fig.~\ref{fig:schematic}(b). In an experiment, for a given material, neither
the strength nor the density of the disorders can be readily adjusted.
However, the sample size can be controlled. Classically, under a vertical
magnetic field, the electrons move along circular trajectories in the domain.
In experiments, for a larger ring sample with constant disorder density, an
electron encounters more disorders/scattering events in one complete rotation.
For computational convenience, we fix the outer radius of the ring domain
to be unity. In this case, varying the disorder density is equivalent to
changing the size of ring domain, where a higher density corresponds to a
larger domain.
Following this heuristic consideration, we fix the
disorder strength as well as the domain size but systematically vary
the density of the disorders. For convenience, in our computations we set
the total number of disorders in the whole domain as a control parameter,
and solve the Dirac equation to obtain the magnitudes of the persistent
currents as a function of the number of disorders. For comparison with
the nonrelativistic quantum counterpart, we solve the Schr\"{o}dinger
equation under the same setting. Our main results are the following.
For the Dirac ring system, as the number of the disorders is systematically
increased, the average current decreases slowly initially and then
plateaus at a finite nonzero value, indicating that the persistent
currents are robust. We demonstrate that WGMs are the physical mechanism
responsible for the robust currents. In contrast, in the nonrelativistic
quantum ring system, the WGMs are sensitive and fragile to the disorders,
leading to a rapid and exponential decay of the currents to zero. We develop
a physical theory based on a quasi one-dimensional approximation to
understand the strikingly contrasting behaviors of the currents in the Dirac
and Schr\"{o}dinger rings. An important implication of our finding
is that persistent currents in the Dirac rings can occur in realistic
systems of large size.

In Sec.~\ref{sec:model}, we describe the Hamiltonian for a 2D Dirac ring
and the numerical method to calculate the persistent currents. In
Sec.~\ref{sec:result}, we demonstrate the robustness of the currents against
random disorders and the emergence of WGMs. In Sec.~\ref{sec:analysis}, we
justify our use of the quasi-1D approximation and derive a physical theory to
understand the drastically contrasting decaying behaviors in Dirac and
Schr\"{o}dinger ring systems. In Sec.~\ref{sec:conclusion}, we present
conclusions and a discussion about the possibility of observing persistent
currents in Dirac systems of large size (e.g., beyond the mesoscopic scale).

\section{Model Hamiltonian and simulation setting} \label{sec:model}

We consider a 2D Dirac ring domain where an AB magnetic flux passes through
the central region, as shown schematically in Fig.~\ref{fig:schematic}(a).
The Dirac Hamiltonian subject to a magnetic field is
\begin{equation} \label{eq:Hamiltonian}
\mathcal{H} = \mathcal{H}_0 + U(\bm{r}) =
\hbar v(\hat{\bm{p}} + e\bm{A})\cdot \hat{\bm{\sigma}}
+ V(\bm{r})\sigma_z + U(\bm{r}),
\end{equation}
where $\hat{\bm{p}}=-i\hbar\partial/\partial\mathbf{r}$ is the momentum
operator, $\hat{\bm{\sigma}}=[\sigma_x,\sigma_y,\sigma_z]^T$ is the ``vector''
of Pauli matrices, $\bm{A}= \Phi \partial(-\ln{|\bm{r}|})/\partial \bm{r}$
is the magnetic vector potential, and $\Phi$ is
the AB magnetic flux. The disorders are modeled as a random electrical
potential function $U(\bm{r})$, and the mass potential that confines the
Dirac particle in the domain is $V(\mathbf{r})$, where $V=\infty$ for
$r<R_1$ or $r>R_2$. In the polar coordinates, the stationary Dirac equation
in the ring domain can be written as (in the units $\hbar=v=1$)
\begin{eqnarray}\label{eq:polar_eq}
\begin{split}
\left( \begin{array}{cc}
  i [U(r,\theta)-\varepsilon] & e^{-i\theta}(\partial_r-\frac{i}{r}\partial_{\theta}+\frac{\Phi/\Phi_0}{r})\\
  e^{i\theta}(\partial_r+\frac{i}{r}\partial_{\theta}-\frac{\Phi/\Phi_0}{r})  & i [U(r,\theta)-\varepsilon]
  \end{array} \right)\\
   \cdot \Psi(r,\theta)=0,
\end{split}
\end{eqnarray}
where $\varepsilon$ denotes the eigenenergy, the relevant lengths are
normalized by the outer radius $R_2$ (e.g., the inner radius becomes
$\xi=R_1/R_2$), and $\Phi_0=h/e$. In the absence of random
disorders, the Dirac equation in the ring domain can be solved analytically
with the following solutions:
\begin{eqnarray} \label{eq:DE_ana_solution}
\Psi(r,\theta) & = & [\psi^-(r,\theta),\psi^+(r,\theta)]^T \\ \nonumber
& = & e^{im\theta}[e^{-i\theta/2}\chi_{\bar{m}}^-(r),ie^{i\theta/2}\chi_{\bar{m}}^+(r)]^T,
\end{eqnarray}
where $\bar{m}=m+\Phi/\Phi_0$ is the effective quantum number of the
angular momentum and $m=\pm1/2,\pm3/2,\pm5/2\cdots$ are the eigenvalues of
the operator $\hat{\mathcal{J}}_z=-i\partial_{\theta}+\sigma_z$.
Differing from the hard potential boundary condition in the Schr\"odinger
system, which is given by $\psi(r=\xi ,\theta)=\psi(r=1,\theta)=0$, the
infinite mass boundary condition in the Dirac system leads to the following
relation between the two components of the spinor wavefunction~\cite{BFM:1987}:
\begin{equation} \label{eq:component_relation}
\psi^+/\psi^- = i \mbox{sgn}[V]e^{i\theta}.
\end{equation}
The radial part of the whole wavefunction can be expressed as a set of
Hankel functions [see Eq.~(\ref{appeq:haknel}) in Appendix A].

Treating the random disorders as perturbations, we have
\begin{equation} \label{eq:Hmatrix_perturbation}
\sum_{i,j}\langle j|\mathcal{H}|i\rangle = \sum_i E^{(0)}_i
+\sum_{i,j}\langle j|U(r,\theta)|i\rangle,
\end{equation}
where $E^{(0)}_i$ and $|i\rangle$'s are the eigenvalues and eigen wavefunctions
of the unperturbed Hamiltonian $H_0=H_{U=0}$, respectively. The energy levels
of the perturbed system can be solved numerically using the Hamiltonian in
Eq.~(\ref{eq:Hmatrix_perturbation}).

The concrete parameter setting in our simulation is the following. We
model the random disorders through a set of uncorrelated Gaussian
potential functions:
\begin{displaymath}
U(r,\theta) = \sum_{s=1}^N U_s(r_s,\theta_s)= \sum_{s=1}^{N}u_s
e^{-\delta r^2/2\sigma^2},
\end{displaymath}
where $s$ and $N$ are the index and the total number of random impurities in
the domain, $\sigma$ and $u_s$ are the size and the potential height of a single
electric impurity, respectively. We set the cutoff radius of any disorder
to be $\delta r \leq 3\sigma$ and the mean radius of the disorders to be
$(R_2-R_1)/20$. The strength of the disorders is randomly chosen from the
interval $[-u_m/2,u_m/2]$, where $u_m$ is determined in terms of the average
spacing of the first ten energy levels, $\Delta E_{10}$.

Note that, our computations are based on the Dirac equation
(not the tight-binding model), so in principle the number of energy levels
is infinite. For high energies, the spacing between two adjacent levels
tends to be uniform. To be concrete, we focus on the low energy regime and
perform detailed computations for 10 representative levels.

For convenience, we use the superscripts ``$D$'' and ``$S$'' to denote the
results for the Dirac and Schr\"{o}dinger ring domains, respectively. Our
computation gives $\Delta E_{10}^{(S)} \approx 10\Delta E_{10}^{(D)}$ for
$\xi=1/2$. The maximum number of disorders is chosen to be $500$
(corresponding to impurity area ratio $S_{dis}/S_{ring}\simeq 0.43$)
to prevent them from covering the ring domain completely.
The single-level persistent current is conventionally defined as~\cite{RV:1993}
\begin{equation} \label{eq:typ_single_PCs}
I_n = \overline{\sqrt{\langle I_n^2(\Phi)\rangle}},
\end{equation}
where $I_n(\Phi)=-\partial E_n(\Phi)/\partial\Phi$ is the flux-dependent
persistent current associated with the $n$th energy level
and ${\langle \cdots} \rangle$ denotes disorder averaging.
The experimentally measurable persistent current is given
by~\cite{CRG:1989,RV:1993}
\begin{equation}\label{eq:typ_PCs}
I^{typ}(\Phi) = {\sqrt{\langle [\sum_{n=1}^{n_{F}} I_n(\Phi)]^2\rangle}},
\end{equation}
where $n_{F}$ is the maximum energy level below the Fermi energy. The average 
persistent currents over magnetic flux is written as $I^{typ}=\overline{I^{typ}(\Phi)}$. 
Generally, the persistent currents in Eqs.~(\ref{eq:typ_single_PCs}) and
(\ref{eq:typ_PCs}) are normalized by the corresponding disorder-free
currents $I_n^{0}$ and $I^{0}$, respectively. Note that the magnitudes of
the edge currents are different, which gives rise to a net current.

\begin{figure}
\centering
\includegraphics[width=\linewidth]{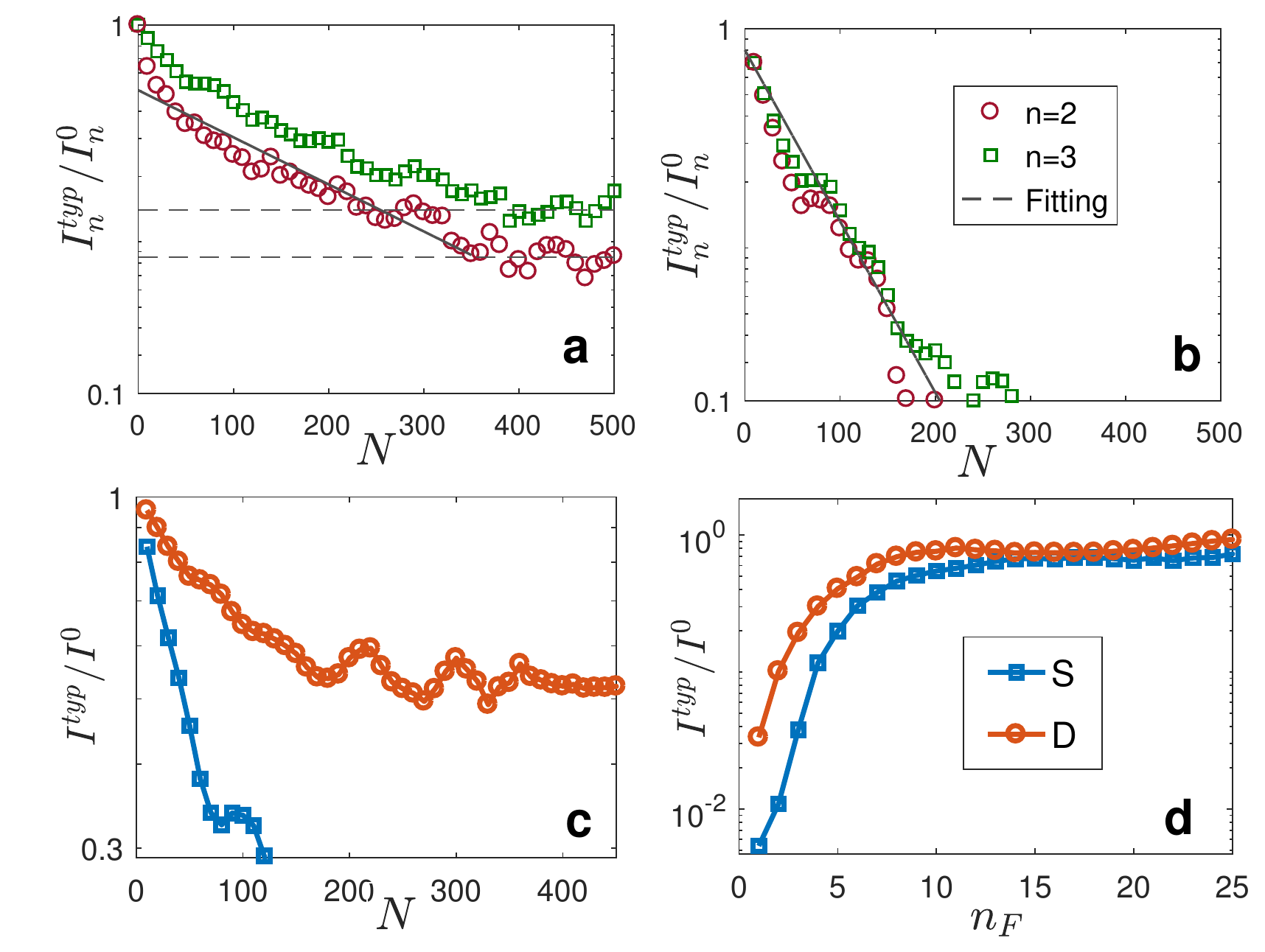}
\caption{ Typical single-level persistent currents versus the
number of the disorders for (a) Dirac and (b) Schr\"{o}dinger rings, for
fixed disorder strength $u_m^{(D,S)}=300\Delta E_{10}^{(D,S)}$. The mean
radius of a single impurity is $\delta r=(1-\xi)/20$. (c) The average
persistent current versus the number of the disorders for five levels
below the Fermi energy, where other parameters are the same as in (a,b) 
for the Dirac (denoted as ``$D$'' and illustrated as circles) and 
Schr\"{o}dinger (denoted as ``$S$'' and displayed as squares) rings. The 
range of the number of disorders, $N\in[0,500]$, corresponds to the range 
of the ratio between the disorder and ring areas $S_{dis}/S_{ring}\in[0,0.43]$. 
(d) Total persistent current versus $n_F$, the number of levels
below the Fermi energy, for the Schr\"{o}dinger (blue squares) and Dirac (red 
circles) ring systems, where the number of impurities is 400.}
\label{fig:Idecay}
\end{figure}

\section{Results and qualitative understanding} \label{sec:result}

Figures~\ref{fig:Idecay}(a) and \ref{fig:Idecay}(b) show, for the Dirac
and Schr\"{o}dinger systems, respectively, the typical single-level
persistent currents versus the number $N$ of random disorders, which are
calculated using $10^2$ statistical realizations. The error in the calculated
value of the current amplitude is about $10^{-2}$. In both cases, the
current amplitude decays exponentially for $0 < N \alt 350$ (i.e.,
$0 < S_{dis}/S_{ring} \alt 0.3$):
$I_n\sim I_n^0$exp$[-\gamma^{(D,S)} N]$, with the distinct feature that the
decay rate for the Dirac system is about half of that for the Schr\"{o}dinger
system: $\gamma^{(D)}/\gamma^{(S)}\approx 1/2$. A more remarkable feature is
that, for the Dirac system, after an initial exponential decay, the current
amplitude approaches a constant value of about $10^{-1}$ (which is about
one order of magnitude larger than the numerical error) for $N \geq 350$, but
for the corresponding Schr\"{o}dinger system the current amplitude
effectively decays to zero. We see that, as more random disorders are
introduced into the domain (or equivalently, as the domain size is increased),
the decaying behavior of the persistent currents is characteristically
different for the Dirac and Schr\"{o}dinger systems: for the former
the currents are robust and continue to exist (in spite of deterioration
in the amplitude) but for the latter the currents quickly diminish. That
is, persistent currents in the Dirac system are robust against random
disorders.
The decay behaviors of the average persistent current associated with the
five lowest energy levels for both the Dirac and Schr\"odinger rings
are demonstrated in Fig.~\ref{fig:Idecay} (c).

The behavior of the total persistent current versus $n_F$, the number
of levels below the Fermi energy, is shown in Fig.~\ref{fig:Idecay}(d), 
where there are 400 random impurities in the ring domain for both the
Schr\"{o}dinger (blue squares) and Dirac (red circles) cases. The quantity
$n_F$ is increased from $1$ to the value when the total persistent
current is saturated. The currents for both systems increase with
$n_F$ for $n_F\in[1,10]$ and become plateaued for further increase in
the value of $n_F$. In the increasing phase, the persistent current of
the Dirac ring system is much larger than that of the Schr\"{o}dinger
counterpart (note the logarithmic scale for the current). That is, for
low energy levels (e.g., $n_F \leq 10$), disorders have a more devastating
effect on the total persistent current for the Schr\"{o}dinger system. However,
the saturated current value for the Dirac ring is not significantly larger
than that for the Schr\"{o}dinger system (again note the logarithmic scale),
due to the reason that the contributions to the total current from higher
energy levels are less sensitive to disorders than those from the low
energy levels.

\begin{figure}
\centering
\includegraphics[width=\linewidth]{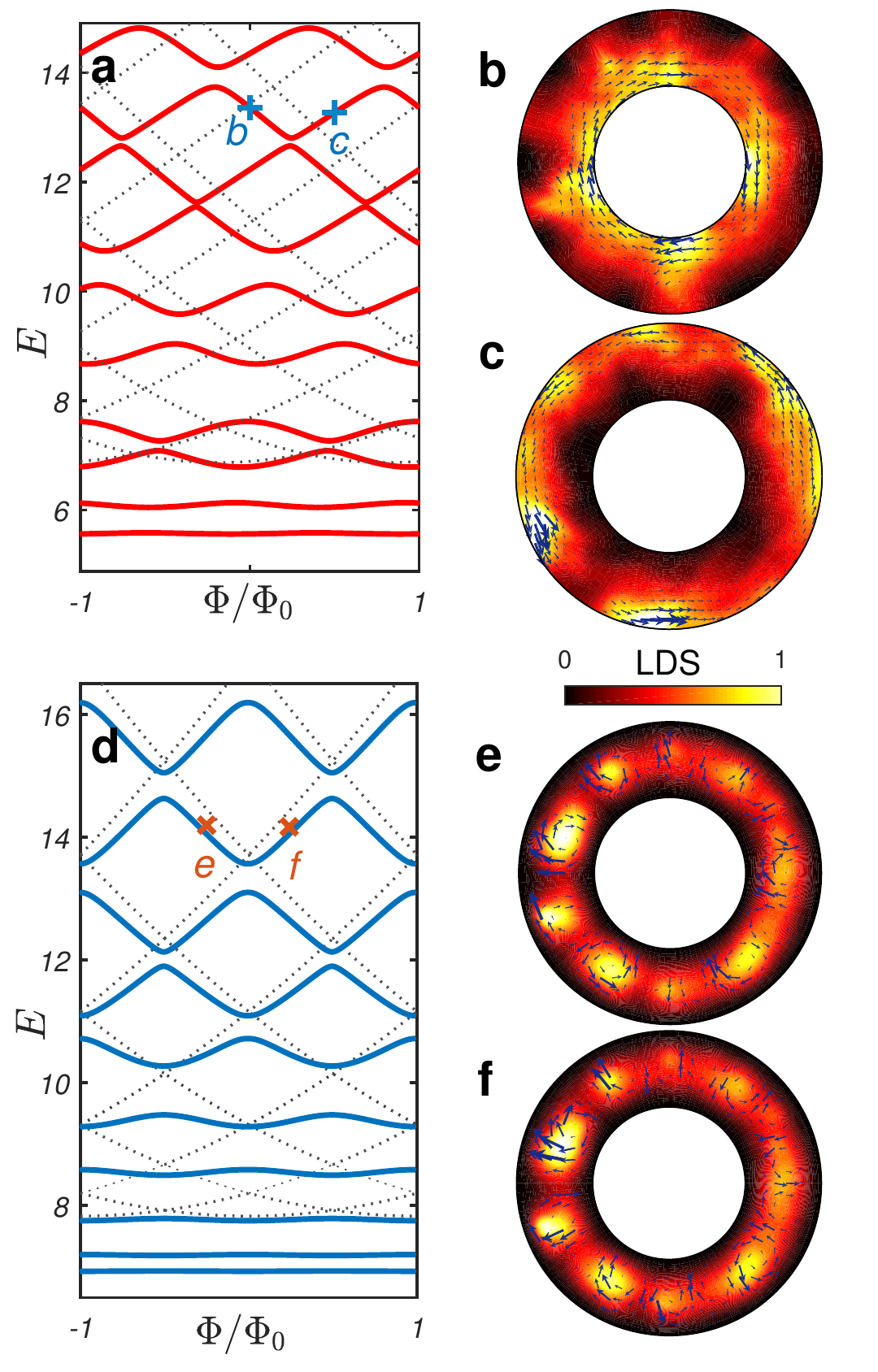}
\caption{(a) Energy level versus the strength of the magnetic flux for
the Dirac system. Gray dashed curves are for the case without random
disorders, and the solid curves represent the energy levels with
100 random disorders (box b and c are for the $9$th energy level in the
disordered system). LDS and LCD patterns are shown for the outer (b) and
inner (c) states. The locations of the states in (b) and (c) are indicated
by the crosses in (a). (d-f) Energy level versus the strength of the
magnetic flux and the LDS and LCD patterns for the Schr\"odinger system. The
locations of the states in (e) and (f) are indicated by the crosses in (d).}
\label{fig:WGM_LDS}
\end{figure}

One question is whether Klein tunneling is responsible for the robust
persistent currents in the Dirac ring system.
In relativistic quantum systems, Klein tunneling is referred to as the
phenomenon where an incoming electron can penetrate a potential barrier
whose height is larger than the electron energy with probability
one~\cite{KNG:2006}. In 1D systems with random impurities, Klein tunneling
has a strong manifestation in the behavior of the persistent
currents~\cite{GS:2014}.
This is because, in one dimension, the incident ``angle'' of a wave
on an impurity is zero so that the condition for Klein tunneling is
always met. However, in two dimensions, there can be a wide
distribution of the incident angle~\cite{KNG:2006} on a random potential,
and the angle range for Klein tunneling is quite limited. In our setting,
the potential field of an electric disorder is Gaussian, rendering highly
unlikely Klein tunneling. To provide further evidence that Klein tunneling
plays little role in sustaining the persistent currents in the Dirac ring
system with disorders, we set the impurity potential to be attractive in
the range $[-u_m,0]$ so that no Klein tunneling can occur. We obtain
essentially the same result as in Fig.~\ref{fig:Idecay}(a).

\begin{figure}
\centering
\includegraphics[width=\linewidth]{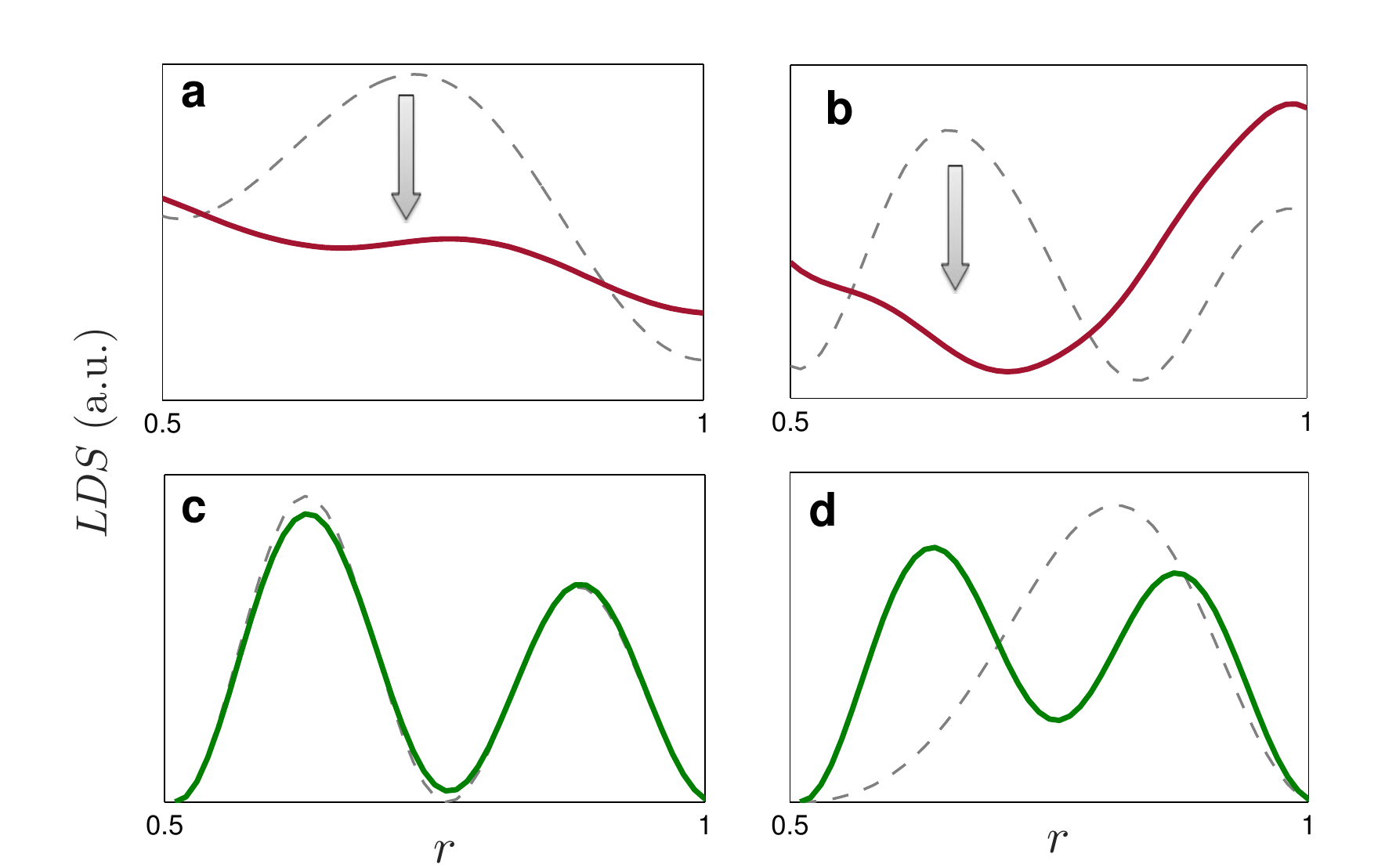}
\caption{(a,b) The radial wavefunctions of the ``clean'' Dirac systems (gray
dashed) of the 22nd and the 25th levels, respectively. The solid curves
in (a,b) show the corresponding average radial wavefunctions
for the case where there are 100 random disorders in the domain.
(c,d) The radial wavefunctions of the Schr\"odinger system with the same
parameters as (a,b). Green solid and gray dashed curves denote the
disordered influenced and the ``clean'' wavefunctions,'' respectively.}
\label{fig:radial_LDS}
\end{figure}

The physical mechanism for persistent currents to sustain in the Dirac
ring system with random disorders can be attributed to a set of WGM states
near the domain boundaries. This can be verified by examining the
local density of states (LDS) and the local current density (LCD) that
can be calculated~\cite{BFM:1987} as
$\bm{j}(r,\theta)=\Psi^{\dag}(r,\theta)\hat{\bm{\sigma}}\Psi(r,\theta)$.
The LDS and LCD distributions are shown in Fig.~\ref{fig:WGM_LDS},
where the WGM characteristics of the boundary states are apparent.
In general, a Dirac fermion tends to stay near one of the infinite mass
boundaries with a high probability, due to the zero-flux boundary
condition. For a Dirac ring domain without any random disorder, the intrinsic
circular symmetry stipulates identical radial wavefunction for different
angular modes, i.e., the Hankel functions with different angular
momentum quantum numbers.
In this case, the WGMs tend to be ``attached'' to the boundaries
of the ferromagnetic material. As a result, there is an asymmetry in the
plot of the energy levels versus the magnetic flux, as shown Fig.~3(a).
Random disorders break the circular symmetry and, as a result,
the WGMs tend to be detached from but they are still near the boundaries.
In general, the LDS and LCD patterns depend on the wavevector and the
detailed distribution of the random disorders. For comparison, we also
plot the LDS and LCD patterns for the Schr\"odinger ring, where the LCD
is calculated by $\bm{j}(r,\theta)=\Psi^{\dag}(r,\theta)(\hat{\bm{\nabla}}
+\bm{A})\Psi(r,\theta)$. As shown in Figs.~\ref{fig:WGM_LDS}(e,f), the LCD
is highly localized due to the disorders.

To further understand the robustness of the WGMs in the Dirac ring domain
against random disorders, we examine the wavefunctions at higher energy levels.
Without any disorder, while the LDS pattern ``attaches'' to the boundary,
its radial wavefunction of high levels ($n\geq9$) is typically maximized
in the interior of the domain, as shown by the dashed black curves in
Fig.~\ref{fig:radial_LDS}. Random disorders attenuate the LDS in the
interior region but its values remain significant near the boundaries,
as shown by the solid curves in Figs.~\ref{fig:radial_LDS}(a,b).
For the Schr\"odinger system, as shown in Figs.~\ref{fig:radial_LDS}(c,d),
the disorders have little effect on the pattern of the average radial
wavefunction. However, the azimuthal components of the wavefunctions are
affected [c.f., the 2D LDS patterns in Figs.~\ref{fig:WGM_LDS}(e,f)].

\begin{figure}
\centering
\includegraphics[width=\linewidth]{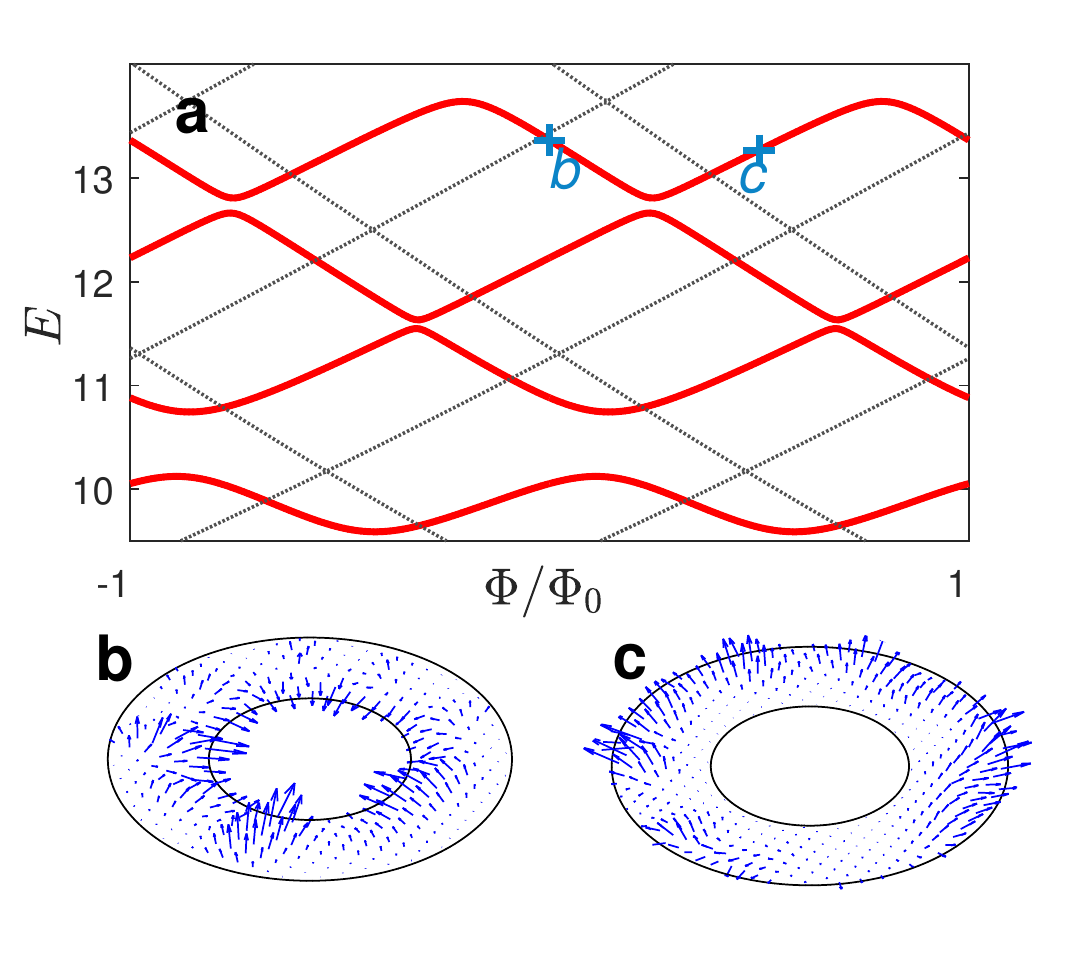}
\caption{For the Dirac ring system, the spin
orientations corresponding to the LDS patterns in
Fig.~\ref{fig:WGM_LDS}(a-c). The locations of the states in (b) and (c)
are indicated by the crosses in (a). Both states correspond
to the 9th level, as in Fig.~3}.
\label{fig:spin}
\end{figure}

A fundamental feature of the Dirac system, which is absent in the
Schr\"{o}dinger counterpart, is the spin texture. We find that the
spin texture associated with the WGMs is hardly affected by the random
disorders. For a 2D Dirac system (e.g., the surface of a 3D topological
insulator), the spin orientation is given by~\cite{FL:2013}
$\bm{s}(r,\theta)=\Psi^{\dag}(r,\theta) \hat{\bm{S}} \Psi(r,\theta)$,
with $\hat{\bm{S}}=(1/2)(\sigma_y,-\sigma_x,\sigma_z)$. As shown
in Fig.~\ref{fig:spin}, the spin orientations of
the WGMs near the inner and outer boundaries tend to be parallel to their
respective normal vectors. (For non-boundary states, the spin orientations
are random.) The robustness of the spin orientation against random
disorders can again be attributed to the zero-flux boundary condition
that allows the states with a definite spin orientation to close on itself
after completing a circular path to ensure constructive interference.
That is, the infinite mass boundaries in the Dirac system tend to ``protect''
the spin orientation for WGM type of boundary states.

\section{Physical understanding of the robust persistent currents in the
Dirac ring} \label{sec:analysis}

For a circular Dirac domain, in the absence of random disorders the energy
level $\varepsilon$ depends on the angular momentum quantum number $m$:
$\varepsilon=\varepsilon(m)$. If the thickness of the quantum
ring is not large, as an approximation~\cite{WFK:1994} we can assume that
the disorders have little effect on the radial component $\chi(r)$ of the
eigenfunctions but they can affect significantly the azimuthal
component $\phi(\theta)$. A general wavefunction from the Dirac equation
can be written as a linear combination of the eigenfunctions:
$\Psi(r,\theta)=\sum_n\phi_n(\theta)\chi_n(r)$, where $\chi_n(r)$ is the
eigenfunctions of the Bessel's differential equation (see
Appendix~\ref{APP:orthogonal_cond} for details), and $\phi_n(\theta)$ is the
azimuthal wavefunction associated with the original angular momentum
quantum number $m$ in the absence of disorders. The orthogonality
condition for $\chi_{n,m}^{\pm}(r)$ is (Appendices A and B)
\begin{displaymath}
\int_\xi^1 dr\frac{1}{r}{\chi_{n^{\prime},m}^{\pm {\ast}}}(r)\chi_{n,m}^{\pm}(r) = \delta_{n^{\prime},n}.
\end{displaymath}
Utilizing this condition and eliminating the radial partial terms in
Eq.~(\ref{eq:polar_eq}), we obtain the governing equation for the
quasi 1D azimuthal wavefunction as
\begin{eqnarray} \label{eq:azimuthal_eq}
&&\partial_{\theta}\phi_{n,m}(\theta) =
\mathcal{\hat{G}}\phi_{n^{\prime},m}(\theta), \\
&&\hat{\mathcal{G}}=i \sum_s \sum_{n^{\prime}}
\left(
  \begin{array}{cc}
  m-1/2 & -e^{-i\theta}\Gamma_{nn^{\prime},m}^{(s)} \\
 -e^{i\theta}\Gamma_{nn^{\prime},m}^{(s)}   & m+1/2
  \end{array} \right),
\end{eqnarray}
where $s$ is the index of the random disorders, $n$ and $n^{\prime}$ are
the energy level indices, $\Gamma_{nn^{\prime},m}^{(s)}$ is the scattering
integral associated with the radial component, which is given by
\begin{equation} \label{eq:scat_int}
\begin{split}
\Gamma_{nn^{\prime},m}^{(s)}(r_s)& =
-\int_{\xi}^1{\chi_{n,m}^{- {\ast}}}(r)U_s(r,\theta)\chi_{n^{\prime},m}^+(r)dr \\
&=\int_{\xi}^1{\chi_{n,m}^{+ {\ast}}}(r)U_s(r,\theta)\chi_{n^{\prime},m}^-(r)dr.
\end{split}
\end{equation}

\begin{figure}
\centering
\includegraphics[width=\linewidth]{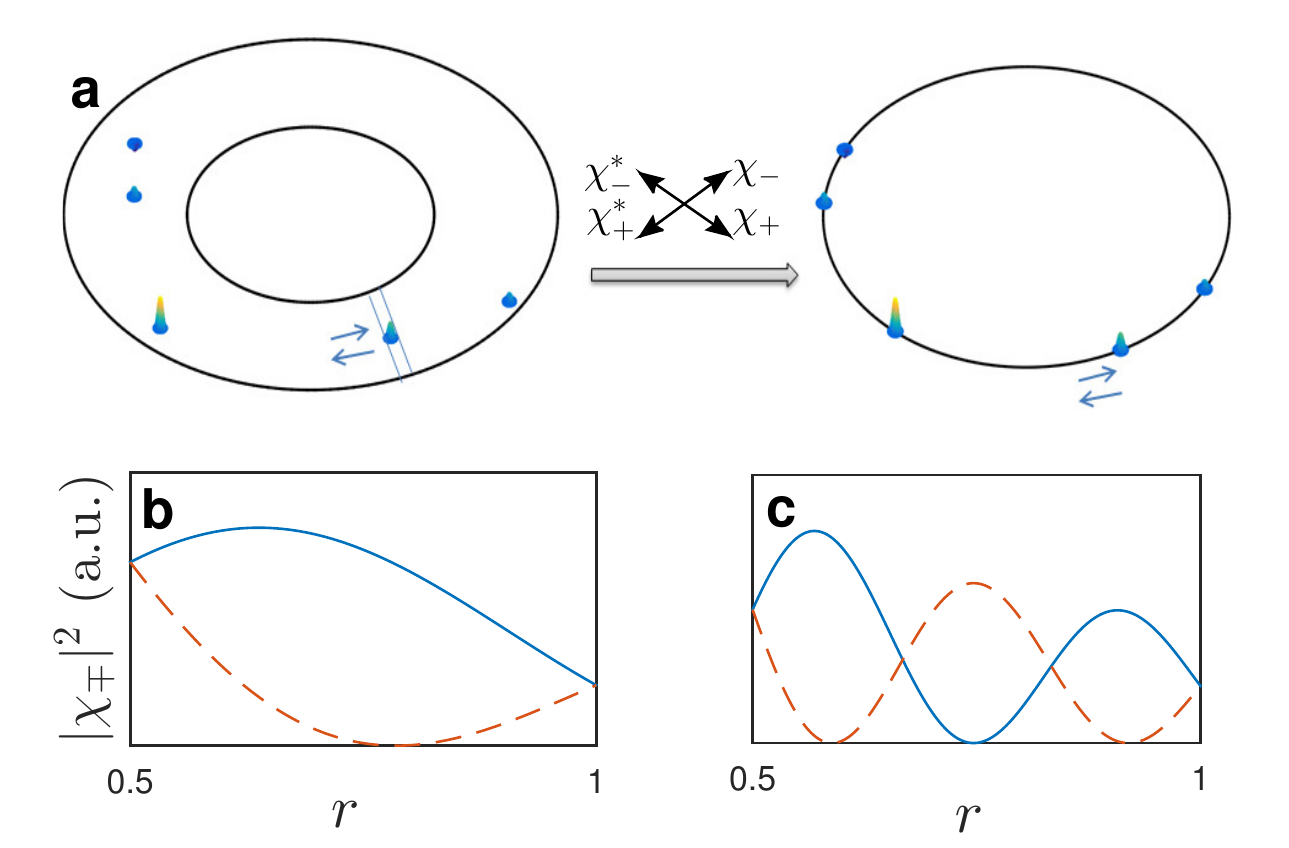}
\caption{(a) Schematic illustration of quasi one-dimensional model
by integrating over the radial dimension. The integral is over a
multiplication of the upper and lower components of the radial
wavefunctions, as indicated in (a). (b,c) The upper (solid) and
lower (dashed) radial wavefunctions for the 11th and 15th energy
levels, respectively.}
\label{fig:Schem_Q1D}
\end{figure}

The approximation procedure is schematically illustrated in
Fig.~\ref{fig:Schem_Q1D}(a), with two representative radial wavefunctions
in \ref{fig:Schem_Q1D}(b) and \ref{fig:Schem_Q1D}(c), respectively. For
simplicity, we set the potential characterizing the random disorders as
\begin{displaymath}
U(r,\theta)=\sum_s U_s(r_s,\theta_s)=\sum_s(u_s/r)\delta(r-r_s)\delta(\theta-\theta_s),
\end{displaymath}
which yields
\begin{displaymath}
\Gamma_{nn^{\prime},m}^{(s)} =
-u_s{\chi_{n,m}^{- {\ast}}}(r_s)\chi_{n^{\prime},m}^+(r_s)\delta(\theta-\theta_0).
\end{displaymath}
For comparison with the nonrelativistic quantum counterpart, we note that
for a Schr\"odinger domain, the azimuthal equation is~\cite{WFK:1994}
\begin{displaymath}
(\partial_{\theta}^2+m^2)\phi_n(\theta) =
\sum_s\sum_{n^{\prime}}\Gamma_{n^{\prime}n,m}^{(s)}\phi_{n^{\prime}}(\theta),
\end{displaymath}
where
\begin{displaymath}
\Gamma_{n^{\prime}n,m}^{(s)} =
u_s\chi_{n^{\prime},m}^{\ast}(r_s)\chi_{n,m}(r_s)\delta(\theta-\theta_s).
\end{displaymath}
Having obtained a quasi 1D equation that approximately describes the
effects of the random impurities, Eq.~\eqref{eq:azimuthal_eq}, we are
in a position to set up a quantum transport model based on the transfer
matrix approach. In particular, the transfer operator $\mathcal{T}$ for
one complete rotation in the ring domain subject to random impurities
is defined as~\cite{BTBB:2007,Titov:2007}
\begin{equation}
\phi(\theta=2\pi)=\mathcal{T}\phi(\theta=0),
\end{equation}
with
\begin{equation} \label{eq:TD_matrix}
\mathcal{T}=\mathcal{T}_P^{(N+1)}  \prod_{s=N}^1
\mathcal{T}_M^{(s)} \mathcal{T}_P^{(s)},
\end{equation}
where the operators $\mathcal{T}_P^{(s)}$ and $\mathcal{T}_M^{(s)}$
represent the propagating and scattering processes. The operators can be
obtained from the first-order Neumann solution~\cite{DAM:1989,DA:1991,R:1993}
of the azimuthal Dirac equation~(\ref{eq:azimuthal_eq}) as
\begin{equation} \label{eq:Dirac_azimuthal}
\phi(\theta_{s^{\prime}})=\hat{Q}
\exp{\left[\int_{\theta_s}^{\theta_{s^{\prime}}}
\hat{\mathcal{G}}(\theta)d\theta\right]}\phi(\theta_s)
\end{equation}
with $\hat{Q}$ denoting the Dyson ordering operator and $\hat{\mathcal{G}}$
being an angle-dependent operator. Analytically it is infeasible
to obtain the solutions of Eq.~\eqref{eq:Dirac_azimuthal}. To gain
insights, we set $\hat{Q}=1$ to obtain the following expressions:
\begin{eqnarray} \label{eq:PM_matrix}
&\mathcal{T}_P^{(s)}=\left(
\begin{array}{cc}
e^{i(\theta_s-\theta_{s-1})(m-1/2)} & 0\\
0 & e^{i(\theta_s-\theta_{s-1})(m+1/2)}
\end{array}
\right), \\
&\mathcal{T}_M^{(s)}=\left(
\begin{array}{cc}
\cos{\Gamma_{nn^{\prime},m}^{(s)}} & -ie^{i\theta}
\sin{\Gamma_{nn^{\prime},m}^{(s)}} \\
-ie^{-i\theta} \sin{\Gamma_{nn^{\prime},m}^{(s)}}
& \cos{\Gamma_{nn^{\prime},m}^{(s)}}
\end{array}
\right).
\end{eqnarray}
Note that these expressions are different than those from the
Schr\"{o}dinger counterpart. To carry out the analysis further, we have
that the transfer matrix associated with the magnetic flux periodicity
for $\theta=0$ is given by
\begin{displaymath}
\mathcal{T}_{\Phi}=e^{i 2\pi\Phi/\Phi_0}\mathcal{I}.
\end{displaymath}
Thus the relationship $m=m(\varepsilon)$ in the presence of random disorders
can be solved when the transfer operator with disorder scattering can match
the transfer matrix associated with the magnetic flux. We have
\begin{equation} \label{eq:detT}
\mathrm{Det}[\mathcal{T}-\mathcal{T}_{\Phi}]=0.
\end{equation}
In our heuristic analysis, we make the diagonal approximation: $n=n^{\prime}$,
so as to avoid generating any additional energy crossings~\cite{WFK:1994}.

\begin{figure}
\centering
\includegraphics[width=\linewidth]{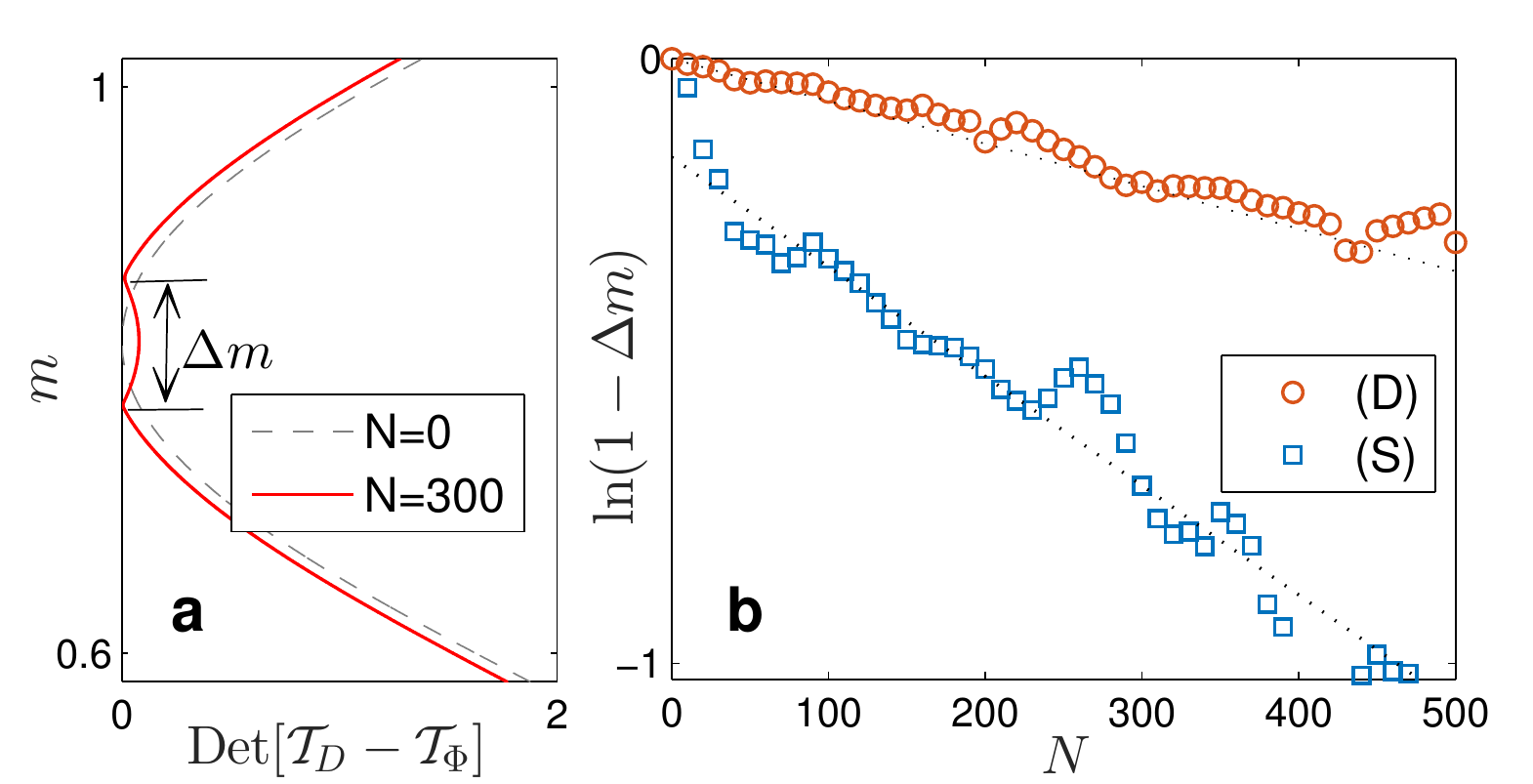}
\caption{ For the Dirac ring domain, (a) the angular momentum
quantum number $m$ versus the value
of the determinant of the transfer matrix, $\mathrm{Det}[T_D-T_{\Phi}]$, for
two cases where there is no disorder (dashed gray curves) and where there are
$300$ disorders (solid curves). (b) Decaying behavior of $1-\Delta m$
with the number $N$ of random disorders for Dirac ($u_m^{(D)}=0.03$) and
Schr\"odinger ($u_m^{(S)}=0.075$) systems on a semi-logarithmic scale.}
\label{fig:Dm}
\end{figure}

Figure~\ref{fig:Dm}(a) shows the relation between the determinant of
the difference of the transfer matrices
$\mathrm{Det}[\mathcal{T}_D - \mathcal{T}_{\Phi}]$ with
the angular momentum quantum number $m$. In the absence of random disorders,
Eq.~(\ref{eq:detT}) has a single solution, which corresponds to the energy
crossing point. With random disorders, energy repulsion occurs, leading to
a split in the original angular momentum quantum number: $\Delta m$, on
which the amplitude of the persistent current depends. Note that the range
of $\Delta m$ is $[0,1]$. For a small value of $\Delta m$, the energy
repulsion is weak so that a large current can be maintained. On the contrary,
for a large value of $\Delta m$ the current amplitude becomes small.
Roughly, the current amplitude is proportional to $1-\Delta m$.
Figure~\ref{fig:Dm}(b) shows the decreasing behavior of $1-\Delta m$
on a semi-logarithmic scale with the number of disorders.
We see that the exponential decay rate is much smaller for
the Dirac system than for the Schr\"odinger counterpart. For a relatively
large value of $N$ (e.g., $N\approx 400$), for the Dirac system the quantity
$1-\Delta m$ tends to plateau at a small (but nonzero) value,
indicating a strong sustainability of the persistent currents against
random disorders. In striking contrast, for the Schr\"odinger system,
the value of $1-\Delta m$ decays rapidly to zero, indicating that
persistent currents in the nonrelativistic quantum ring are vulnerable to
random disorders. These results are consistent with those from direct
numerical calculations (Fig.~\ref{fig:Idecay}).

Our analysis based on the quasi-1D equation provides a heuristic method
to estimate the decay rate of the persistent currents as the number $N$ (or
density) of random disorders is increased. For an initial range of $N$,
for both the Dirac and Schr\"odinger systems, the decaying behavior of the
currents can be written as $I_n/I_n^0=A$exp$(-\gamma N)$, where the decay rate
is $\gamma\sim\langle\Gamma_{n}\rangle$, where $\Gamma_{n}$ is the
overlapping integral of the radial wavefunctions. Thus the ratio of the decay
rates between the Dirac and Schr\"odinger systems is approximately given by
\begin{equation} \label{eq:decay_ration}
\frac{\gamma_{n}^{(D)}}{\gamma_{n}^{(S)}}\approx
\frac{\langle\Gamma_{n}^{(D)}\rangle}{\langle\Gamma_{n}^{(S)}\rangle},
\end{equation}
where
\begin{eqnarray}
\nonumber
\langle \Gamma_{n}^{(D)} \rangle & = &
\int^1_{\xi}dr{\chi_{n}^{- {\ast}}}\chi_{n}^{+}, \\ \nonumber
\langle \Gamma_{n}^{(S)} \rangle & = &
\int^1_{\xi}drr{\chi_n^{(S) {\ast}}}\chi_n^{(S)}.
\end{eqnarray}
In the Dirac ring, the upper and lower components of the radial
wavefunction have a large phase difference for low energy levels,
as shown in Figs.~\ref{fig:Schem_Q1D}(b,c). As a
result, we have
\begin{displaymath}
\langle \Gamma_{n}^{(D)}\rangle < \langle \Gamma_{n}^{(S)}\rangle.
\end{displaymath}
For example, for the $2$nd and $3$rd energy levels, we have
\begin{displaymath}
{\gamma_{2,3}^{(D)}}/{\gamma_{2,3}^{(S)}} \approx \langle \Gamma_{2,3}^{(D)}
\rangle/\langle \Gamma_{2,3}^{(S)}\rangle \approx 1/2,
\end{displaymath}
which agrees approximately with the numerical results in
Figs.~\ref{fig:Idecay}(c,d).
Note that this disorder resistant scattering mechanism may be less effective
for high energy levels because the phase difference between the upper and
lower components of the corresponding wavefunctions can be negligibly small.
Consequently, the integral in Eq.~(\ref{eq:scat_int}) assumes values
comparable to those in the Schr\"odinger ring system. The implication is
that persistent currents associated with high energy levels in the Dirac
ring are expected not to be robust.

\section{Conclusion and discussion} \label{sec:conclusion}

For a ring domain with a magnetic flux through the center, persistent
currents can arise due to the AB effect. This paper investigates the
effect of random disorders on persistent currents in relativistic quantum
(Dirac) ring systems. There are two reasons to investigate the impact of
random disorders. Firstly, in any realistic materials random disorders
are inevitable. In nonrelativistic quantum systems the disorders have a
devastating effect on the persistent currents, so they can only be
observed in systems of sufficiently small size (e.g., size $\lesssim$ the
phase coherence length of the material). For relativistic quantum systems,
there is a recent work on the effects of random disorders on persistent
currents in one dimension~\cite{GS:2014}. It is of interest to understand
the effect in experimentally more feasible 2D systems. Secondly, in order to
assess the feasibility of observing persistent currents in large systems,
one can study the impact of random disorders of a systematically increasing
density, because to solve the Dirac equation under a magnetic field in
a large system can be computationally demanding. These points can be
elaborated through the following discussion of the main results of this
paper and their implications.

Previous theoretical and experimental results on persistent currents in
nonrelativistic quantum (Schr\"{o}dinger) systems revealed that the currents
are sensitive and thus vulnerable to disorders. A natural question is then
whether persistent currents can be more ``sustainable'' in relativistic
quantum systems. Through direct numerical calculation of persistent
currents for both Dirac and Schr\"{o}dinger ring systems with systematically
varying number (or density) of random disorders, we find that the
currents in the Dirac system are significantly more robust than
those in the Schr\"{o}dinger counterpart. While for both systems, as the
number of random disorders is increased from zero, the current amplitude
decays exponentially, there are two key characteristic differences between
the relativistic and nonrelativistic quantum cases. Firstly, the rate of decay
is much smaller in the Dirac than in the Schr\"{o}dinger system. Secondly,
for the Dirac ring the exponential decay is terminated when the number
of random disorders becomes large and is subsequently replaced by a
plateaued behavior with a finite current amplitude, but in the
Schr\"{o}dinger ring the exponential decay continues until the currents
effectively become zero. The underlying quantum states providing
``sustained'' persistent currents in the Dirac system are found to be
a WGM type of boundary states. We developed a physical theory, based on
a quasi 1D approximation, to explain the distinct current decaying behaviors
in the Dirac and Schr\"{o}dinger systems. Specifically, under this approximation
the effect of random disorders can be assessed and the persistent currents
can be calculated through a scattering integral over the radial dimension
that involves the product of the two components of the relativistic quantum
spinor. These findings suggest the extraordinary robustness of persistent
currents in the Dirac system, due to the robustness of the underpinnings
of the currents, the WGM states, to random disorders.

Our calculations uncovered that, for both the Dirac and Schr\"{o}dinger rings,
the interior states are vulnerable to random disorders. It is the zero-flux
boundary condition that renders the WGM boundary states robust in the Dirac
system. (In the Schr\"{o}dinger system boundary states cannot form due to
the Dirichlet boundary condition.) It is possible to observe the sustaining
boundary states experimentally by exploiting, e.g., the surface states of
a 3D topological insulators, where a ring domain can be formed through
the deposition of ferromagnetic insulating materials on the surface of the
topological insulator. Another finding is that the spin orientations of
the WGM states are hardly affected by random disorders, which may have
implications for relativistic quantum spintronic devices.

For the Dirac and Schr\"odinger systems, the energy dispersion relation is
linear and parabolic, i.e., $E=k$ and $E=k^2$ (with a proper normalization),
respectively. Under the normalization the lowest energy level is larger
than unity for both cases. For the same Fermi energy, the wavelength in the
Dirac ring is smaller than that for Schr\"odinger counterpart. The
robust persistent currents in the Dirac ring thus are not an effect of a
larger wavelength and a weaker sensitivity to disorder. In fact, the
robust persistent currents are due to the whispering gallery modes along
the edges, as stipulated by the zero-flux boundary conditions in the Dirac
ring.

An important implication of our finding lies in the possibility to observe
persistent currents in Dirac systems of large sizes.
In Schr\"{o}dinger materials (normal metals or semiconductors), the currents
can be observed but only when the device size is smaller than or close to
the phase coherence length so that the electron trajectories are ballistic
or approximately ballistic with short diffusion time. When the device size
is much larger than this scale, random scattering will be strong, diminishing
the circulating current.
However, the robustness of the persistent currents in the Dirac system implies
that the relativistic quantum phenomenon can occur in larger devices,
possibly on the macroscopic scale. This can be argued by noting that,
as the ring size is increased, the number of scattering events
that a particle experiences in one circulating motion will increase.
From the standpoint of scattering, increasing the density of random
disorders for fixed device size is equivalent to enlarging the device.
For strong random disorders, Anderson localization sets in~\cite{Anderson:1958},
prohibiting currents inside the domain. However, because of the strong
boundary currents in Dirac fermion systems, it is possible that the
persistent currents will not vanish. Below we provide an estimate of
the maximally possible system size in which persistent currents can sustain.

In experimental studies, a 2D Dirac ring can be realized through the
surface states of, e.g., Bi$_2$Te$_3$/Bi$_2$Se$_3$, with Fermi velocity about
$v_F\approx7\times10^5\mbox{m}/\mbox{s}$~\cite{ZLQDFZ:2010,Xia:2009,
BRSGDHCY:2011}. In our simulation, the Gaussian-like disorder is analogous to
charge puddles of size $\sim 30$nm and strength $\sim 10$meV associated with
the surface states of Mn/Ca-doped Bi$_2$Te$_3$/Bi$_2$Se$_3$
materials~\cite{BRSGDHCY:2011}. In a pure Bi$_2$Te$_3$/Bi$_2$Se$_3$ sample,
the strength of the charge puddles is smaller than that for doped materials.
We can thus set $u_m/2=5$meV. In our computation, for the case of high
disorder density, say $400-500$ impurities, the disorder pattern
is quite similar to that of the charge puddles from
experiments~\cite{BRSGDHCY:2011}. The maximum strength of the charge puddles
is given by $u_m=300\Delta E_{10}$, with
$\Delta E_{10}=\hbar v_F \Delta k_{10}/R_2$, where $k_{10}R_2=0.45$ and
$R_2$ is the outer radius of the ring, which can be
estimated as $R_2=300\times0.45\hbar v_F u_m\approx 6\mu\mbox{m}$.
As a result, the estimated size of the Dirac ring in which robust
persistent currents can exist is $D=2R_2\sim 12\mu\mbox{m}$, which is much
larger than the maximum size of the normal metallic or semiconductor rings
with persistent currents observed in previous experimental
studies~\cite{LDDB:1990,CWBKGK:1991,MCB:1993,RSMHE:2001,KBFGG:2007,
BSPGVGH:2009,BKBHM:2009,CNSJH:2013}.

In a clean Dirac ring of size $D=12 \mu\mbox{m}$, the persistent currents
associated with one energy level can be estimated as
$I_n^0\sim 2\Delta E_{10}/\Phi_0=0.45\times\hbar v_F/(R_2\Phi_0)\approx3$nA,
where $\Phi_0=h/e\sim4\times10^{-15}\mbox{Tm}^2$ is the magnetic flux quantum.
Even if there are $500$ impurities in the ring domain, there is still
a finite persistent current: $I_n\approx 0.1I_n^0\approx0.3$nA. The total
persistent current in an experimental system is given by
$I=\sum_{n=1}^{N}I_n$, where the integer $N$ depends on the Fermi energy.
For example, if the Fermi energy is $E=1$meV, several energy levels will
be included. The total persistent current is $I\sim1nA$, which can be
observed in experiments, e.g., by using the SQUID technique~\cite{LDDB:1990,
CWBKGK:1991,MCB:1993,RSMHE:2001,BSPGVGH:2009,BKBHM:2009}.

\section*{Acknowledgement}

We thank Dr.~H.-Y. Xu for valuable discussions.
We would like to acknowledge support from the Vannevar Bush
Faculty Fellowship program sponsored by the Basic Research Office of
the Assistant Secretary of Defense for Research and Engineering and
funded by the Office of Naval Research through Grant No.~N00014-16-1-2828.
This work was also supported by ONR under Grant No.~N00014-15-1-2405.

\appendix
\appendixpageoff

\section{Orthonormality of radial wavefunctions} \label{APP:orthogonal_cond}

The radial component of a Dirac spinor in 2D is governed by
\begin{eqnarray} \label{appeq:radial_eq}
\left(
\begin{array}{cc}
 0&\frac{d}{dr}+\frac{\bar{m}+1/2}{r}      \\
   -\frac{d}{dr}+\frac{\bar{m}-1/2}{r} & 0
\end{array}
\right)\chi=i\varepsilon\chi.
\end{eqnarray}
The two decoupled equations for the upper and lower components of the
radial wavefunction can be written as $H_r^{\prime}\chi=0$, where
\begin{equation}\label{appeq:dirac_bessel}
\begin{split}
\left[\frac{d^2}{dr^2}+\frac{1}{r}\frac{d}{dr} +
\left(\varepsilon^2-\frac{(\bar{m}-1/2)^2}{r^2}\right) \right]\chi_{n,m}^-=0, \\
\left[\frac{d^2}{dr^2}+\frac{1}{r}\frac{d}{dr} +
\left(\varepsilon^2-\frac{(\bar{m}+1/2)^2}{r^2}\right)\right]\chi_{n,m}^+=0.
\end{split}
\end{equation}
The solutions of these equations can be expressed in terms of the
Hankel functions:
\begin{eqnarray}\label{appeq:haknel}
\left(
\begin{array}{c}
 \chi_m^-    \\
  \chi_m^+
\end{array}
\right)=
\frac{1}{\sqrt{N}}\left(
\begin{array}{c}
H_{\bar{m}-1/2}^{(1)}(\varepsilon r) +
\alpha H_{\bar{m}-1/2}^{(2)}(\varepsilon r) \\
iH_{\bar{m}+1/2}^{(1)}(\varepsilon r) +
i\alpha H_{\bar{m}+1/2}^{(2)}(\varepsilon r)
\end{array}
\right),
\end{eqnarray}
where the coefficient $\alpha$ and the normalized coefficient $N$ are given by
\begin{equation} \label{appeq:coefficient}
\begin{split}
\alpha \ &=-\frac{H_{\bar{m}+1/2}^{1}(\varepsilon\xi) +
H_{\bar{m}-1/2}^{1}(\varepsilon\xi)}
{H_{\bar{m}+1/2}^{2}(\varepsilon\xi)+H_{\bar{m}-1/2}^{2}(\varepsilon\xi)} \\
&=-\frac{H_{\bar{m}+1/2}^{1}(\varepsilon)-H_{\bar{m}-1/2}^{1}(\varepsilon)}
{H_{\bar{m}+1/2}^{2}(\varepsilon)-H_{\bar{m}-1/2}^{2}(\varepsilon)}, \\
N_m &=2\pi\int_{\xi}^{1}rdr(|{\chi_m^-}^{\prime}|^2+|{\chi_m^+}^{\prime}|^2),
\end{split}
\end{equation}
respectively, with $\chi_{1,2}^{\prime}$ denoting the unnormalized radial
wavefunctions. Consider two different pairs of quantum numbers:
${m_i,\varepsilon_i}$ and ${m_j,\varepsilon_j}$, where $i\neq j$. Following a similar derivation method in Ref.~\cite{WFK:1994} and
using Eq.~(\ref{appeq:dirac_bessel}), we
obtain~\cite{WFK:1994}
\begin{equation}
\begin{split}
(m_j^2&-m_i^2)\int_{\xi}^{1}\frac{dr}{r}
{\chi_{m_j}^{\pm {\ast}}}(r)\chi_{m_i}^{\pm}(r,\varepsilon_i) \\
&=(\varepsilon_j^2-\varepsilon_i^2) \int_{\xi}^{1}rdr
{\chi_{m_j}^{\pm {\ast}}}(r,\varepsilon_j)\chi_{m_i}^{\pm}(r,\varepsilon_i).
\end{split}
\end{equation}
Setting $\varepsilon_i=\varepsilon_j=\varepsilon$, we have
$m_{i,j}=m_{i,j}(\varepsilon)$ and
\begin{equation}
\begin{split}
[m_j^2(\varepsilon)-m_i^2(\varepsilon)]&\int_{\xi}^{1}\frac{dr}{r}
{\chi_{m_j(\varepsilon)}^{\pm {\ast}}}(r,\varepsilon)
\chi_{m_i(\varepsilon)}^{\pm}(r,\varepsilon)=0.
\end{split}
\end{equation}
For nondegenerate energy levels, if $m_i(\varepsilon)\neq m_j(\varepsilon)$,
the integral with the weight $1/r$ is zero. For
$m_i(\varepsilon)=m_j(\varepsilon)$, the integral can assume an arbitrary
value and, for convenience, we can set it to be unity. As a result, the
orthonormal condition becomes
\begin{equation} \label{appeq:new_orthonormality}
\int_{\xi}^{1}\frac{dr}{r}{\chi_{m_j(\varepsilon)}^{\pm {\ast}}}(r,\varepsilon)
\chi_{m_i(\varepsilon)}^{\pm}(r,\varepsilon)=\delta_{i,j},
\end{equation}
leading to the normalized condition
\begin{equation}
{N^{\pm}_{m}}^{\prime}=2\pi\int_{\xi}^{1}\frac{dr}{r}
|{\chi_m^{\pm}}^{\prime}|^2.
\end{equation}

\section{Azimuthal equation with random disorders in the Dirac ring
system}

Substituting the entire wavefunction into the Dirac equation in the
polar coordinates with random disorders [Eq.~(\ref{eq:polar_eq})], we
have the equations for the upper and lower components of the spinor as
\begin{widetext}
\begin{equation}
\begin{split}
\phi_{n,m}^+\left(\partial_r+\frac{\Phi/\Phi_0}{r}\right)\chi_{n,m}^+
-\chi_{n,m}^+\frac{i}{r}\partial_{\theta}\phi_{n,m}^+
+ie^{i\theta}\phi_{n,m}^-(U_s-\varepsilon)\chi_{n,m}^-=0, \\
\phi_{n,m}^-\left(\partial_r-\frac{\Phi/\Phi_0}{r}\right)\chi_{n,m}^-
+\chi_{n,m}^-\frac{i}{r}\partial_{\theta}\phi_{n,m}^-
+ie^{-i\theta}\phi_{n,m}^+(U_s-\varepsilon)\chi_{n,m}^+=0.
\end{split}
\end{equation}
\end{widetext}
Since Eq.~(\ref{appeq:radial_eq}) can be expressed as
\begin{widetext}
\begin{equation}
\left(\partial_r+\frac{\Phi/\Phi_0}{r}\right)\chi_{n,m}^+
=-\frac{m+1/2}{r}\chi_{n,m}^+ +i\varepsilon\chi_{n,m}^- \ \ \mbox{and} \ \
\left(\partial_r-\frac{\Phi/\Phi_0}{r}\right)\chi_{n,m}^-
= \ \  \frac{m-1/2}{r}\chi_{n,m}^- +i\varepsilon\chi_{n,m}^+,
\end{equation}
\end{widetext}
we can eliminate the term in the radial dimension: $\partial_r\chi^{\pm}$.
In particular, making the approximation
$e^{i\theta}\phi_{n}^-\approx\phi_{n}^+$,
we can express the azimuthal equation in the matrix form
\begin{widetext}
\begin{eqnarray}\label{appeq:whole_eq}
\sum_n\frac{1}{r} & \left(
\begin{array}{cc}
\partial_{\theta}-i(m-1/2) & 0 \\
0 & \partial_{\theta}-i(m+1/2)
\end{array}
\right)
\left(
\begin{array}{cc}
\phi_{n}^-\chi_{n,m}^-  \\
\phi_{n}^+\chi_{n,m}^+
\end{array}
\right)
=\sum_n\left(
\begin{array}{cc}
0 & e^{-i\theta}U_s(r,\theta) \\
e^{i\theta}U_s(r,\theta) & 0
\end{array}
\right)
\left(
\begin{array}{cc}
\phi_{n}^-\chi_{n,m}^-  \\
\phi_{n}^+\chi_{n,m}^+
\end{array}
\right).
\end{eqnarray}
\end{widetext}
Multiplying
$[{\chi_{ n^{\prime},m}^{-{\ast}}}, {\chi_{n^{\prime},m}^{+{\ast}}}]$
on both sides of Eq.~(\ref{appeq:whole_eq}), integrating over $r$ in
the region $[\xi,1]$, and using the orthonormal condition in
Eq.~(\ref{appeq:new_orthonormality}), we can simplify the azimuthal
equation for the Dirac system as
\begin{widetext}
\begin{equation}
\left(
  \begin{array}{cc}
  \partial_{\theta}-i(m-1/2) & 0\\
 0  & \partial_{\theta}+i(m+1/2)
  \end{array}
   \right)
\phi(\theta)
=\sum_s\sum_{n^{\prime}}
\left(
  \begin{array}{cc}
  0 & -e^{-i\theta}\int_{\xi}^{1}dr{\chi_{n^{\prime},m}^{- {\ast}}}^U_s(r,\theta)
\chi_{n,m}^+\\
 e^{i\theta}\int_{\xi}^1dr{\chi_{n^{\prime},m}^{+ {\ast}}}U_s(r,\theta)\chi_{n,m}^-  & 0
  \end{array}
   \right)
   \phi.
\end{equation}
\end{widetext}

\section{Scattering matrix method for the Schr\"odinger system}
\label{APP:SM_Schr}

Based on the same approximation as for the  Dirac system, we have
the azimuthal equation for the Schr\"odinger case as~\cite{WFK:1994}
\begin{equation}
(\partial_{\theta}^2+m^2)\phi_n(\theta) =
\sum_s\sum_{n^{\prime}}\Gamma_{n^{\prime}n,m}^{(s)}\phi_{n^{\prime}}(\theta),
\end{equation}
where $\Gamma_{n^{\prime}n,m}^{(s)}=u_s\chi_{n^{\prime},m}^{\ast}(r_s)
\chi_{n,m}(r_s)\delta(\theta-\theta_s)$. The orthonormal condition
is~\cite{WFK:1994}
\begin{displaymath}
\int_{\xi}^{1}dr (1/r) \chi_{m_j(\varepsilon)}^{\dag}(r,\varepsilon)
\chi_{m_i(\varepsilon)}(r,\varepsilon)=\delta_{i,j}.
\end{displaymath}
The azimuthal wavefunction of the Schr\"odinger system with an impulsive
impurity satisfies the boundary conditions
\begin{equation}\label{appeq:BC_schr}
\begin{split}
\phi_n(\theta_s^+)&=\phi_n(\theta_s^-), \\ \\
\frac{d\phi_n(\theta)}{d\theta} |_{\theta=\theta_s^+}
-\frac{d\phi_n(\theta)}{d\theta}|_{\theta=\theta_s^-}
&=\sum_{n^{\prime}}\Gamma_{n^{\prime}n,m}^{(s)}\phi_{n^{\prime}}(\theta).
\end{split}
\end{equation}
Similar to the Dirac system, we make the diagonal approximation:
$n=n^{\prime}$. To avoid numerical divergence, we use the scattering
matrix method. In particular, for propagation along a free path and
scattering from an impurity, the respective scattering matrices can be
obtained from Eq.~(\ref{appeq:BC_schr}):
\begin{eqnarray}
&&\mathcal{S}_P^{(s)}=\left(
\begin{array}{cc}
 0 & e^{i(\theta_s-\theta_{s-1})m} \\
 e^{i(\theta_s-\theta_{s-1})m} & 0
\end{array}
\right), \\
&&\mathcal{S}_M^{(s)} =
\left(
\begin{array}{cc}
 -\frac{i\Gamma_{nn,m}^{(s)}e^{2im}}{2m-i\Gamma_{nn,m}^{(s)}} &
\frac{2m}{2m-i\Gamma_{nn,m}^{(s)}} \\
\frac{2m}{2m-i\Gamma_{nn,m}^{(s)}}  &
-\frac{i\Gamma_{n,n}^{(s)}e^{-2im}}{2m-i\Gamma_{nn,m}^{(s)}}
\end{array}
\right)
\end{eqnarray}
The total scattering matrix is given by
\begin{equation}
\mathcal{S} = \mathcal{S}_P^{(N+1)} \otimes \mathcal{S}_M^{(N)}
\otimes \mathcal{S}_P^{(N)} \otimes  \cdots \otimes
\mathcal{S}_M^{(1)} \otimes \mathcal{S}_P^{(1)}.
\end{equation}
If we consider two scattering matrices defined by
\begin{equation}
\mathcal{S}_{i}=\left(
\begin{array}{cc}
 r_i & t_i^{\prime} \\
  t_i & r _i^{\prime}
\end{array}
\right),  \ \ \
\mathcal{S}_{j}=\left(
\begin{array}{cc}
 r_j & t_j^{\prime} \\
  t_j & r _j^{\prime}
\end{array}
\right).
\end{equation}
The compounded scattering matrix
$\mathcal{S}_{ij}=\mathcal{S}_{i}\otimes \mathcal{S}_{j}$
can be calculated as~\cite{BOOK:datta,TA:1991,FM:2011}
\begin{equation}
\mathcal{S}_{ij}=\left(
\begin{array}{cc}
  r_i+t_i^{\prime}r_j(1-r_i^{\prime}r_j)^{-1}t_i  &
t_i^{\prime}(1-r_jr_i^{\prime})^{-1}t_i \\
 t_j(1-r_i^{\prime}r_j)^{-1}t_i  &
r_j^{\prime}+t_jr_i^{\prime}(1-r_jr_i^{\prime})^{-1}t_j^{\prime}
\end{array}
\right).
\end{equation}
Combining the total scattering matrix for a set of random disorders with
the scattering matrix associated with the magnetic flux
\begin{equation}
\mathcal{S}_{\Phi}=\left(
\begin{array}{cc}
0                    & e^{-i2\pi\Phi/\Phi_0}   \\
e^{i2\pi\Phi/\Phi_0} & 0
\end{array}
\right),
\end{equation}
we have~\cite{TA:1991,FM:2011}
\begin{equation}
\mathrm{Det}[\mathcal{S}-\mathcal{S}_{\Phi}]=0,
\end{equation}
from which the angular momentum quantum number $m$ and its split value
$\Delta m$ can be solved.


%
\end{document}